\documentclass[smallextended]{svjour3}       
\smartqed  
\usepackage{graphicx}
\begin{document}

\title{A decaying factor accounts for contained activity in neuronal networks with no need of hierarchical or modular organization
}

\titlerunning{A decaying factor accounts for contained activity in neuronal networks}        

\author{
Diego Raphael Amancio \and
Osvaldo Novais Oliveira Jr. \and
Luciano da Fontoura Costa 
}


\institute{Institute of Physics of S\~ao Carlos \at
              University of S\~ao Paulo, P. O. Box 369, Postal Code 13560-970 \\
              S\~ao Carlos, S\~ao Paulo, Brazil\\
              \email{diego.amancio@usp.br, diegoraphael@gmail.com}           
}

\date{Received: date / Accepted: date}

\maketitle

\begin{abstract}
The mechanisms responsible for contention of activity in systems represented by networks are crucial in various phenomena, as in diseases such as epilepsy that affects the neuronal networks, and for information dissemination in social networks. The first models to account for contained activity included triggering and inhibition processes, but they cannot be applied to social networks where inhibition is clearly absent. A recent model showed that contained activity can be achieved with no need of inhibition processes provided that the network is subdivided in modules (communities). In this paper, we introduce a new concept inspired in the Hebbian theory through which activity contention is reached by incorporating a dynamics based on a decaying activity in a random walk mechanism preferential to the node activity. Upon selecting the decay coefficient within a proper range, we observed sustained activity in all the networks tested, viz. random, Barabasi-Albert and geographical networks. The generality of this finding was confirmed by showing that modularity is no longer needed if the dynamics based on the integrate-and-fire dynamics incorporated the decay factor. Taken together, these results provide a proof of principle that persistent, restrained network activation might occur in the absence of any particular topological structure. This may be the reason why neuronal activity does not outspread to the entire neuronal network, even when no special topological organization exists.

\end{abstract}

\section{Introduction}
\label{intro}

The human brain is an example of complex system~\cite{csystem,csystem2,csystem3}, where the local dynamics is unable to predict the global functionality because the brain components relate to each other in a non-trivial way. This is why the brain has often been represented as complex networks~\cite{newman,survey} at various scales~\cite{cit2,cit3}, ranging from neurons linked by synapses to interconnected large brain regions~\cite{watts98,sporns2004}. With this type of representation, one may learn about brain functionality and operation. For example, the network framework allowed the observation of synchronization patterns and the establishment of conditions for epileptic seizures, in terms of the topological organization~\cite{cit3,info,percha}. It has been suggested that the human brain works in a critical region analogous to a phase transition dynamics between ordered and chaotic behaviors~\cite{kaiser2}. This critical operation, characterized by a persistent dynamics where the activation neither quickly dies out nor activates the whole network, is essential for studying functional patterns in complex neural networks~\cite{kaiser2,kaiser}. Because the critical operation is responsible for the maintenance of brain functions~\cite{kaiser2} (e.g., in the critical operation processing capabilities are enhanced~\cite{enhanced1,enhanced2} and epileptic seizures are prevented~\cite{kaiser}, it is crucial to determine the factors leading to such restrictions of activity.

While traditional neural networks models require excitatory and inhibitory connections to restrain activity in the critical operation~\cite{curmod1,curmod2}, recent models have shown it to be possible to restrain activity without inhibitory connections~\cite{kaiser2,kaiser}. More specifically, with nodes representing cortical columns~\cite{corticalcolumns} rather than individual neurons, a simple dynamics could explain restrained activity in hierarchical modular neural networks~\cite{bizenger,nature2011,nature2008} under a wide range of initial parameters~\cite{kaiser}. In this paper, we introduce a dynamics that generalizes the approach reported in Ref.~\cite{kaiser}. Taking as inspiration the Hebbian theory, according to which neuronal links are reinforced whenever they are stimulated, we adopted a random walk dynamics preferential to the node activity. We show that the neuronal activity can be topologically constrained even in non-modular structures such as random, geographical and scale-free networks. Therefore, the existence of hierarchy and communities is no longer required. In addition, differently from the model in Ref.~\cite{kaiser}, the number of active nodes in the steady state is weakly dependent on the initial localization of the random walker.

\section{Description of the Preferential Random Walk} \label{descricao}

    Random walks have been employed in a wide variety of contexts~\cite{newmanbook,wang,newrw1,newrw2,teago}, including brain networks as a complementary strategy to the analysis of synchronization properties provided by spectral analysis~\cite{bocateli,estrada,spectra}. Traditional random walks were used to analyze the interplay between structure and function in cortical and neural networks~\cite{wang,corrs}, while self-avoiding random walks~\cite{sarw} were employed to show that the human cortical network is most resilient to brain injuries~\cite{csystem}.  The proposed model of preferential random walk is based on the Hebbian theory~\cite{hebe}, which imposes a rule to determine how the weights connecting neurons in neural networks are updated. Formally, the weight $w_{ij}$ linking neurons $i$ and $j$ is updated according to:
    \begin{equation}
        w_{ij} = p^{-1} \sum_{k=1}^p x_i^k x_j^k,
    \end{equation}
    where $x_i^k$ is the $k$-th input for neuron $i$ and $p$ is the number of training patterns in the artificial neural network~\cite{ann}. Thus, neurons that are repeatedly fired at the same time will tend to have their links strengthened. This effect is simulated through the introduction of a preferential random walk where the particle performing the random walk is more likely to leap onto a node that has already been frequently accessed in the last iterations. Thus, analogously to the Hebbian theory for edges weights, nodes frequently accessed will tend to increase its activity due to the reinforcement effect. The proposed random walk is preferential to the activity $\mathcal{T}$ of each node at the time step $t$. At $t = t_0 = 1$, every node $i$ has activity $\mathcal{T}_i^{t_0} = 1$. Assuming that the particle performing the random walk over the network is at node $i$, having $j$ as its neighbor, the next node to be visited at $t + 1$ is chosen according to the probability $P^{t+1}_{i \rightarrow j}$:
    \begin{equation}
        P^{t+1}_{i \rightarrow j} = \frac{ \mathcal{T}_{j}^t }{ \sum_k a_{ik} \mathcal{T}_{k}^t },
    \end{equation}
    where
    \begin{equation}\label{eq.mavg}
    a_{ik} = \left\{
    \begin{array}{ll}
        1  & \textrm{ if  $i$ and $k$ are connected}, \\
        0  & \textrm{ otherwise.} \\
    \end{array}
    \right.
    \end{equation}
    In other words, the particle tends to propagate toward the most active neighbor. The next node visited will have its activity increased by one unity. At the beginning of every time step $t$, all nodes have their activity decaying according to the following rule:
    \begin{equation} \label{decaimentoeq}
        \mathcal{T}_i^{t+1} = \alpha \mathcal{T}_i^{t},
    \end{equation}
    where $\alpha$ is the rate of activity preservation, which is restricted to the interval $0 < \alpha < 1$. In the experiments, a given node $i$ was considered activated when $\mathcal{T}_i \geq \tau_c$. Arbitrarily, we adopted $\tau_c = 2.5$.

\section*{Results and Discussion}

\begin{figure*}
    \begin{center}
        \includegraphics[width=1\textwidth]{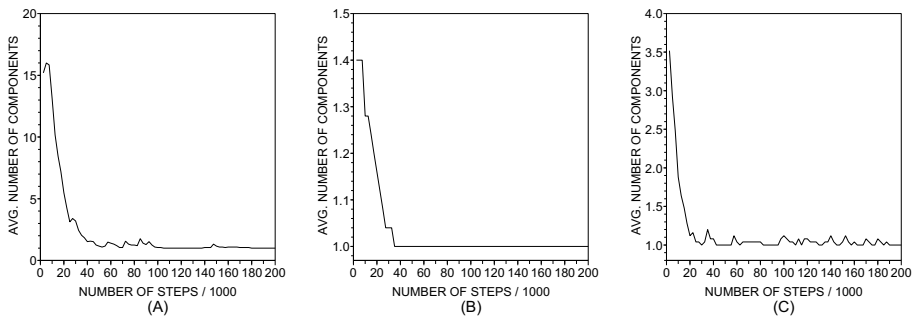}
    \end{center}
    \caption{\label{fig1} Number of components of activated nodes when $t > t_f$ for the (a) random network, built from the Erdos-Renyi model~\cite{randomico}; (b) geographical network, built from a Voronoi diagram~\cite{survey,delaunay}; and (c) scale-free (BA) network, obtained from the Barabasi-Albert model~\cite{sfn,barabasi} {(see degree distributions in Figure S1 of the Supplementary Information)}. The average degree $\langle k \rangle$ and the number of nodes $N$ were fixed in $\langle k \rangle = 3$ and $N = 3000$. All results were obtained by performing 50 simulations. The dynamics was applied for the three networks and the initial node was randomly chosen in each one of the executions. Note that, for the three networks, the dynamics converges to a single component, which characterizes a sustained activation.}
\end{figure*}

Sustained activity was observed with the preferential random walk process provided that the preservation coefficient $\alpha$ (equation \ref{decaimentoeq}) was $\alpha \leq 0.9999$. Figure \ref{fig1} shows that after an abrupt onset of activity in the first few time steps, a sharp decrease occurs for all three networks leading eventually to only a single active connected component\footnote{A component is a subnetwok in which there is a path between all nodes belonging to the component.}. The sharpest decay occurred for the geographical network, since at $t = 4.0 \times 10^4$ activity was restrained to only one component. The activation in geographic networks probably stabilizes faster because these networks are devoid of long-range connections. As such, the activation does not spread to other regions of the network.

Similar results were obtained for the number of active nodes {(Figure S2 of the SI)}, since {after $2.0 \times 10^5$ steps} activity decayed to a small number of nodes for the $3$ networks. Here, however, important differences were noted in comparing the networks. The maximum number of activated nodes was ca. 7\% for the random and BA networks, but for the geographical network this was ca. 1\%. This difference may be ascribed to the small-world property inherent in both BA and random networks. As one should expect, the number of active nodes increases with the preservation factor $\alpha$, as shown in {Figure S3 of the SI}.

\begin{figure*}
    \begin{center}
        \fbox{\includegraphics[width=1\textwidth]{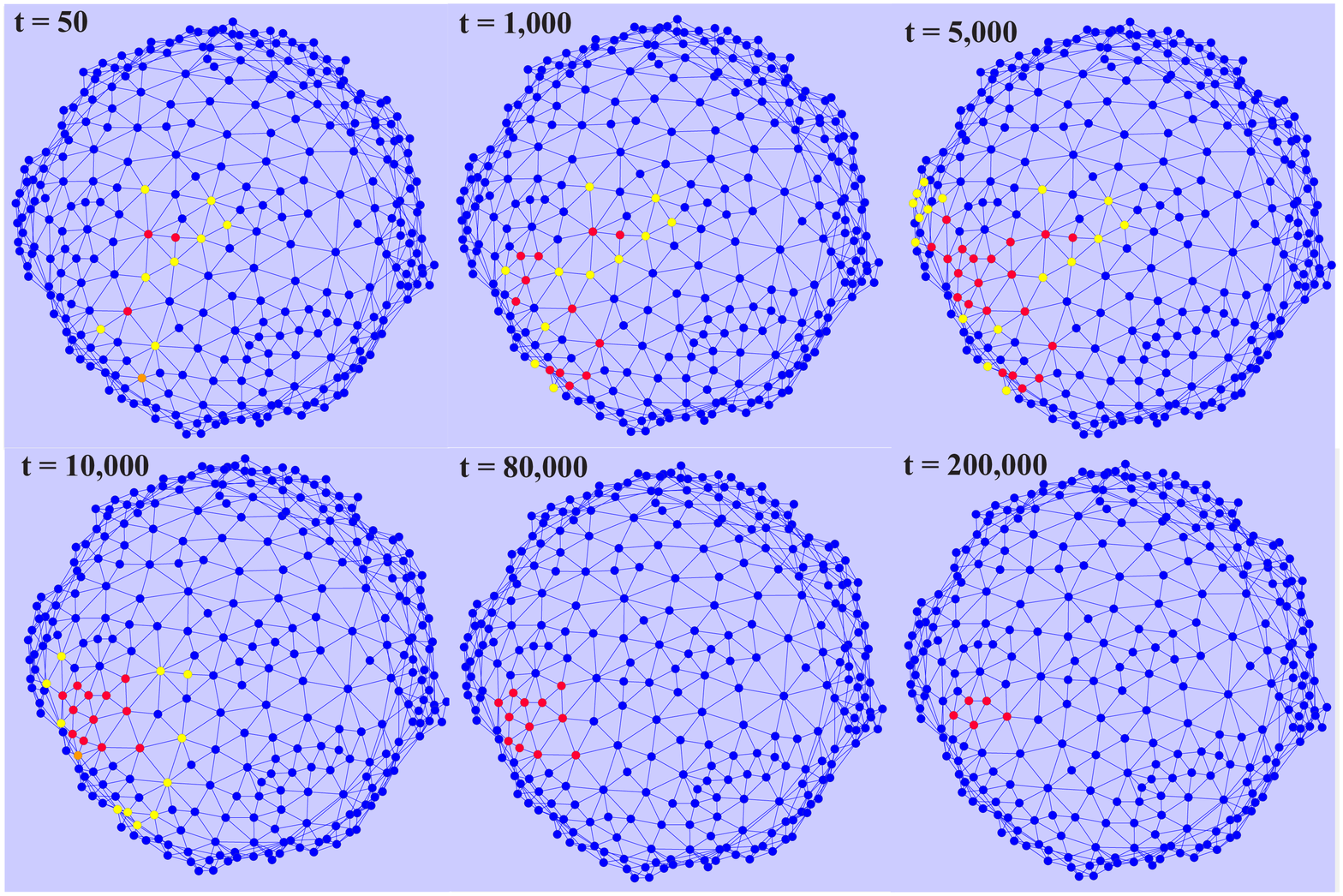}}
    \end{center}
    \caption{\label{fig2.aa} Visualization of activity spreading in a small geographical network with $N = 300$ and $\langle k \rangle = 3$. The colors represent the different scales of activity $\mathcal{T}$: blue, for $1.0 \leq \mathcal{T} < 1.5$; yellow, for $1.5 \leq \mathcal{T} < 2.0$; orange, for $2.0 \leq \mathcal{T} < 2.5$ and red, for $\mathcal{T} \geq 2.5$. Note that in the first steps the activation spreads quickly. Then, the set of active nodes becomes increasingly smaller until reaching the steady configuration at $t = t_f = 200,000$. }
\end{figure*}

\begin{figure*}
        \begin{center}
            \fbox{\includegraphics[width=0.5\textwidth]{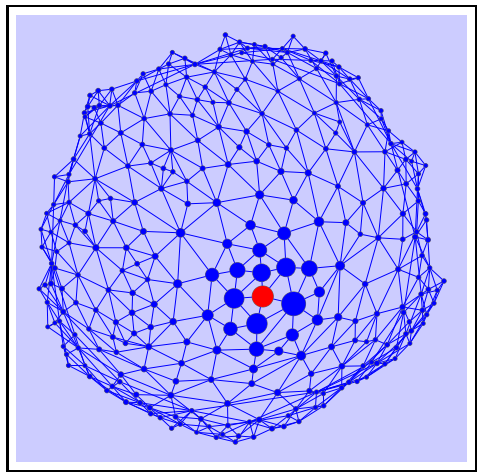}}
        \end{center}
        \caption{\label{fig2.a} Geographical network and the frequency of activation in $300$ executions. The diameter of nodes is proportional to the number of times that a given node was active at $t = t_f$. All the preferential random walks started at the red node.}
\end{figure*}

Further analysis can be made by considering which nodes remain active throughout the simulations. Figure \ref{fig2.aa} shows the evolution of activated nodes in a small geographical network in a given execution of the model (a similar evolution was observed for the other network models). It is worth noting that even when the random walk starts from the same initial node, the set of active nodes at $t=t_f$ is not always the same. Actually, most of the times the final activation is spread along the nearest neighbors from the starting node, as illustrated in Figure \ref{fig2.a} (see also Figure S4 of the SI). The amount of active nodes appears to be independent of the starting node for the random walk, as very little impact was observed on the spreading {(see Figure S5 of the SI)}. This is in contrast with the results in Ref.~\cite{kaiser}.

\begin{figure*}
    \begin{center}
        \includegraphics[width=1\textwidth]{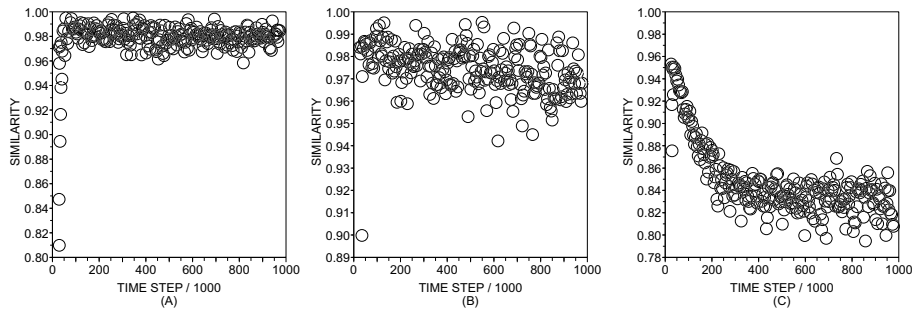}
    \end{center}
    \caption{\label{presRate} Preservation rate $\tau$ of the set of active nodes. $\tau$ is defined as $\tau = | S_{t_o} \cap S_{t_o+\Delta} | / | S_{t_o} |$, where $S_{t_o}$ represents the set of active nodes at $t=t_o$  and $S_{t_o+\Delta}$ represents the set of active vertices at $t = t_o + \Delta$. In this figure, $\Delta = 1,000$. After the first steps, there is almost no variation in the set of active nodes for both random and geographical networks. This means that the set of active nodes remains static. In the case of scale-free network a very different behavior emerged. In the early stages there is a slight variation whereas in the subsequent steps about 83~\% of the set of active nodes vary within $\Delta = 1,000$ steps.}
\end{figure*}

The set of active nodes was largely preserved throughout the time evolution for the random and the geographical networks, as shown in Figure \ref{presRate}. In contrast, only 83\% of the active nodes preserved their activity for the BA network, probably because activity was directed toward the hubs.

\begin{figure*}
    \begin{center}
        \includegraphics[width=1\textwidth]{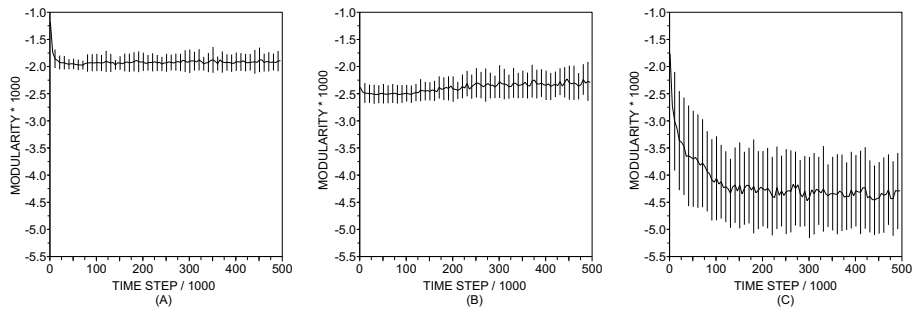}
    \end{center}
    \caption{\label{figmodularity} Evolution of the modularity $Q$ for (a) the random network; (b) the geographical network; and (c) the scale-free network.
    $Q$ is given by $Q = (1/4m) \sum_{ij} ( a_{ij} - k_i k_j / 2m )(1 - s_i s_j)$, where $a_{ij}$ is an element of the adjacency matrix, $k_i = \sum_{i} a_{ij}$ is the degree of node $i$ and $s_i = 1$ if $i$ is an active node and $s_i = 0$ otherwise. While the modularity of the geographical network remains nearly constant, it decreases at the first steps and remains constant for both random and scale-free networks.}
\end{figure*}

An important difference was observed with regard to the results reported by~\cite{kaiser}. Rather than being organized in communities~\cite{nature2011} (i.e., in groups of nodes strongly connected to each other and weakly connected to nodes of other groups), the active nodes were not restrained to communities, as indicated by the degree of modularity $Q$ in Figure \ref{figmodularity} (the higher the value of $Q$, the greater the tendency to restrain activity in a community). The BA network again behaved differently.

The importance of introducing a decay factor in the dynamics was verified by considering the widely used integrate-and-fire dynamics~\cite{ifire}. The latter can be described mathematically using:
\begin{equation}
    T_i^{t+1} = \left\{
    \begin{array}{ll}
        T_i^t + \gamma \sum_{j=1}^N \sum_f a_{ij} \delta(t-t_j^f)  & \textrm{if } T_i^t < \tau_f \\
        \gamma \sum_{j=1}^N \sum_f a_{ij}                           & \textrm{if } T_i^t \geq \tau_f \\
    \end{array}
    \right.
\end{equation}
where $T_i^t$ represents the accumulated potential of node $i$ at time $t$, $t_j^f$ is the time when the $f$-th spike occurs and $\gamma$ is the coupling factor. When the potential $T_i$ reaches a given threshold $\tau_f$, the node sends the accumulated activation throughout the outcome links and its internal state is cleaned. In the simulations we considered $\gamma = 1$ and $\tau_f = 3$. The effect of decay was taken into account simply by multiplying the current potential $T_i^t$ by $\alpha$.

\begin{figure*}
    \begin{center}
        \includegraphics[width=1\textwidth]{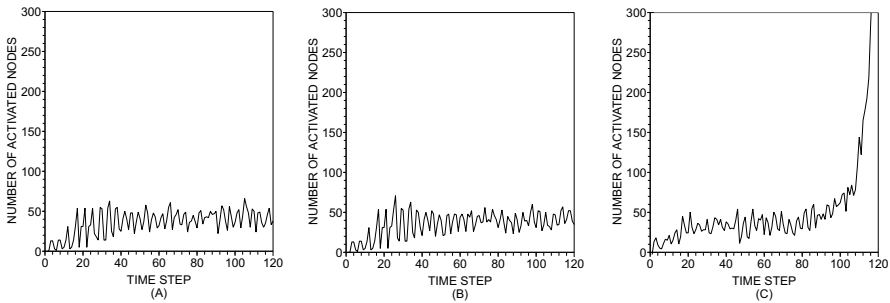}
    \end{center}
    \caption{\label{randd} Dynamics of the set of activated nodes (i.e., nodes with $T_i > \tau_c$) in the random network when (a) $\alpha = 0.9000$, (b) $\alpha = 0.9020$ and $\alpha = 0.9022$. While in (a) in (b) the activity is restrained, in (c) it spreads all over the network.}
\end{figure*}

Figure \ref{randd} confirms that activity is sustained for the random network, provided a proper value for the rate of activity preservation is used. Interestingly, the same conclusion was found for the other two network models {(see Figures S6 and S7 of the SI)}. Therefore, the observation of contained activity for modular networks reported by~\cite{kaiser} no longer applies when the proposed dynamics is adopted.

\section{Conclusion}

The study of the brain with the representation provided by the network framework is one of the most promising issues in neuroscience. In the current paper we have employed complex networks to study the conditions influencing the operation in the critical point. Upon introducing a novel random walk preferential to the activity of nodes (i.e., preferential to the recent frequency of access) inspired in the biological model of Hebb~\cite{hebe}, we found as a proof of principle that persistent and contained network activation can occur not only in the absence of inhibitory nodes, but also in the absence of a hierarchical clustered organization. In particular, this result can be seen as a generalization of previous findings with similar dynamics~\cite{kaiser} advocating that small-worlds and random networks could not restrict the activity without the presence of inhibitory connections. The detailed analysis of the proposed preferential random walk revealed different behaviors for the three networks studied. While the number of active nodes at the steady state is similar both in random and scale-free networks, this quantity is much smaller in geographic networks. This result suggests that the spread is facilitated by the long-range connections and the small-world effect present in such networks. At the steady state, the size of the active set was found to be similar for all three networks. The dependence of the size of this active core on the activity decay was also verified. In all cases, the number of active nodes increases with the preservation factor. We also found that during the first time steps of the spreading, the active nodes move toward the formation of an anti-community structure~\cite{autovetornewman}, in opposition to the behavior observed in other similar dynamics. Finally, we have investigated the influence of the choice of the initial node on some dynamical aspects. More specifically, we found that there is no apparent relationship between the degree of the node locating the random walker at $t=t_0$ and the size of the active core at the steady state. On the other hand, we verified that the choice of the initial node interferes with the localization of the active nodes at $t = t_f$, for the final set of actives nodes spreads along the nearest neighbors from the initial activation.

As future work, the approach may be extended by considering other phenomena along the dynamics. For example, one could examine if a limited sustained activity is kept when the exhaustion phenomena~\cite{kaiser} is taken into account, as some neurological systems do not maintain activation for long periods in order to save brain resources. Also, one could probe the effect from different types of network connectivity on the spreading of activation, as diseases such as epilepsy might be related to distinct connectivity patterns.


\begin{acknowledgements}
	DRA (2010/00927-9) and LFC (2011/50761-2) acknowledge FAPESP for the financial support. The financial support from CNPq is also acknowledged.

\end{acknowledgements}

\bibliographystyle{spbasic}      


\setcounter{table}{0}
\renewcommand{\thetable}{S\arabic{table}}
\renewcommand\tablename{{\bf Table}}

\setcounter{figure}{0}
\renewcommand{\thefigure}{S\arabic{figure}}
\renewcommand\figurename{{\bf Figure}}

\newpage

\section*{Supplementary Information}

\begin{figure}[h]
        \begin{center}
            \includegraphics[width=1\textwidth]{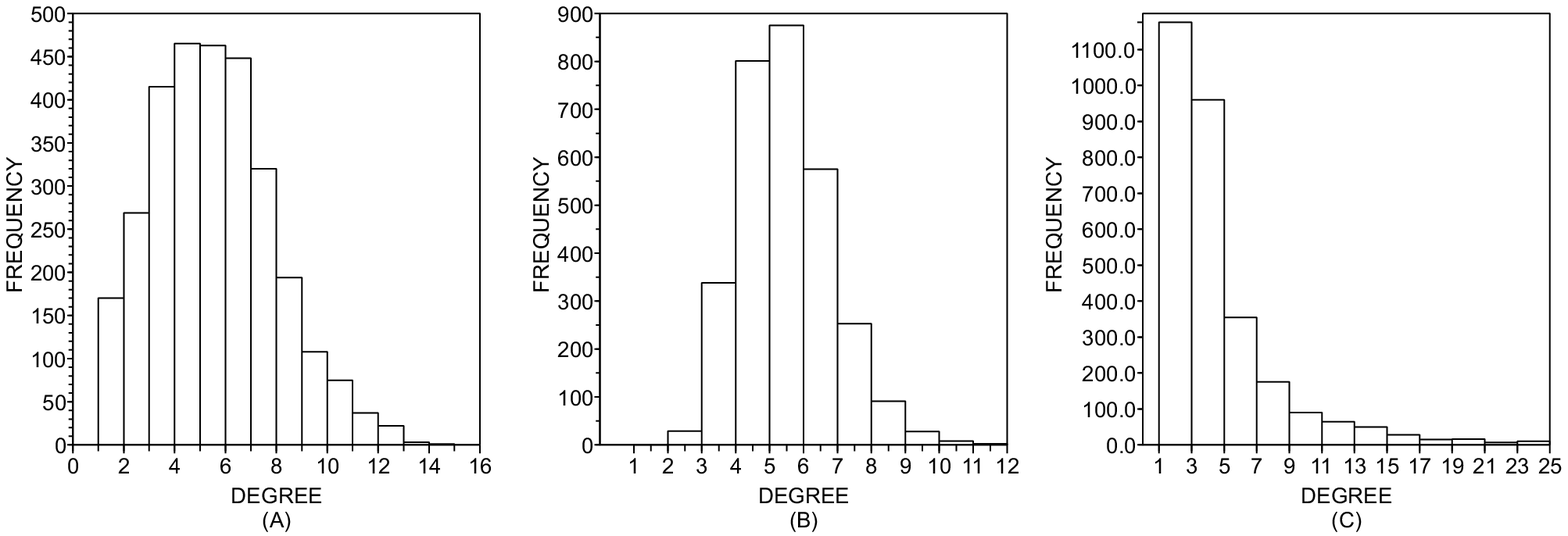}
        \end{center}
        \caption{\label{fig4} Degree distribution for the (a) random network; (b) geographical network; and (c) scale-free network.}
    \end{figure}

    \begin{figure}[h]
        \begin{center}
            \includegraphics[width=1\textwidth]{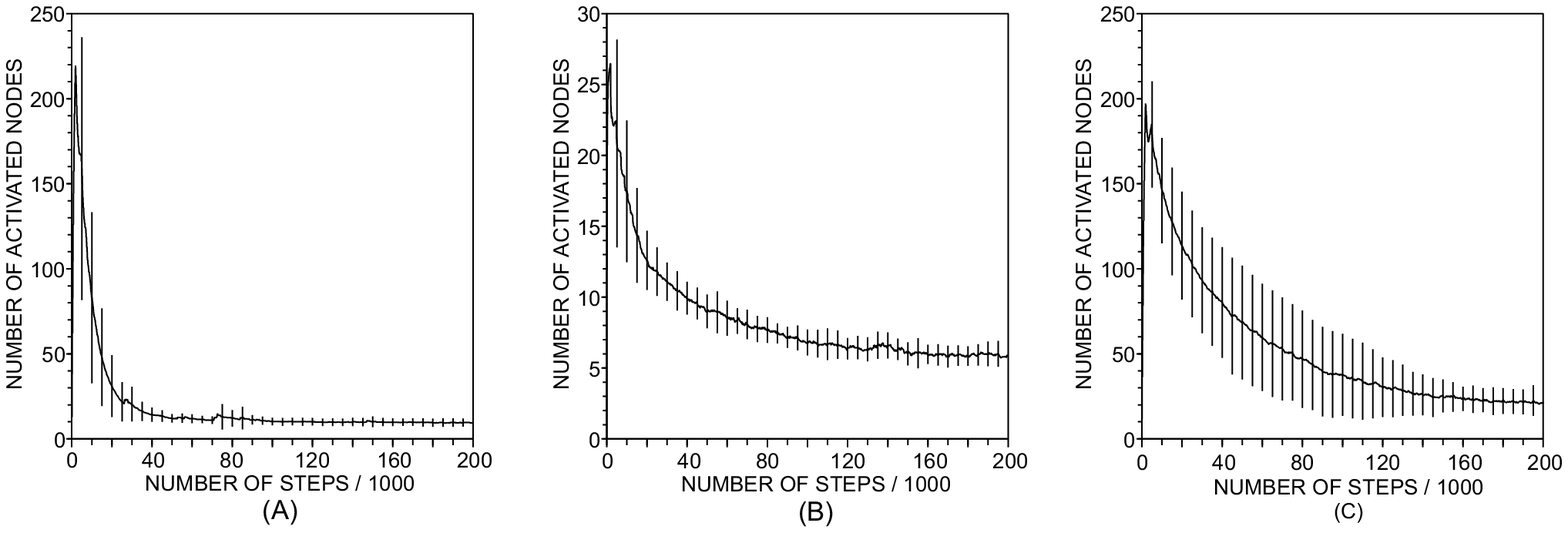}
        \end{center}
        \caption{\label{fig2mod} Number of activated nodes when $t > t_f$ for the (a) random network; (b) geographical network; and (c) scale-free network. In all networks, the dynamics converges to a steady set of activated nodes.}
    \end{figure}

    \begin{figure}[h]
        \begin{center}
            \includegraphics[width=1\textwidth]{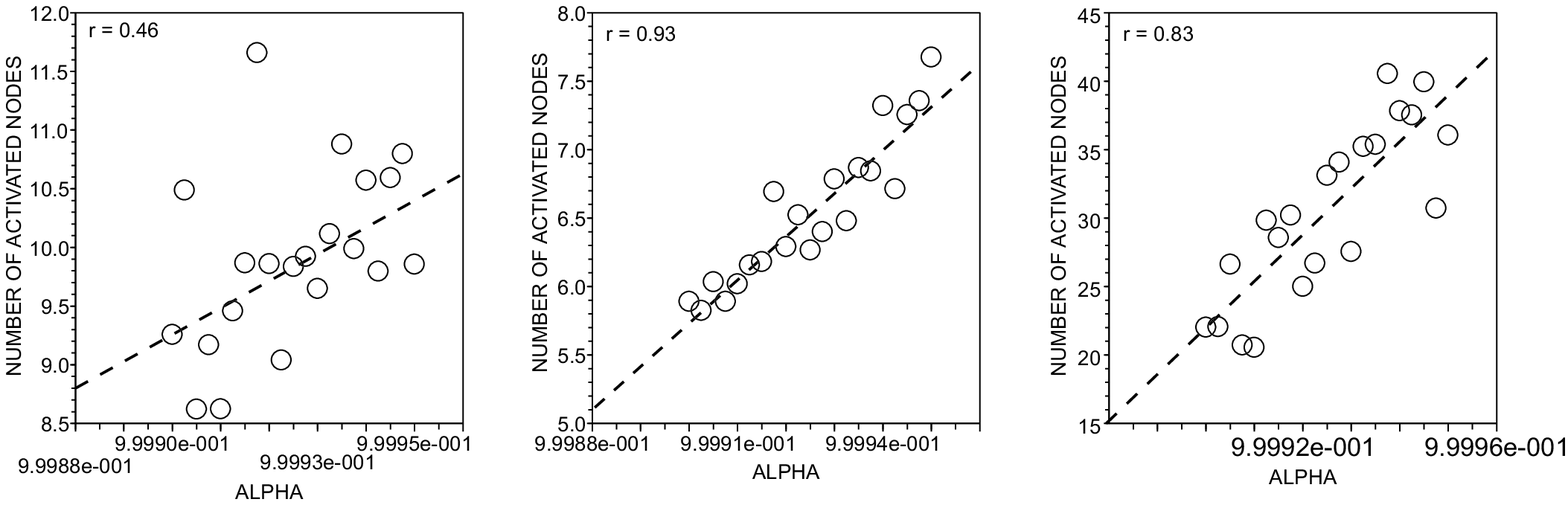}
        \end{center}
        \caption{\label{fig3} Relationship between the rate of activity decay and the number of activated nodes for (a) the random network; (b) the geographical networks; and for (c) scale-free network.
        The sole parameter present in the dynamics of the preferential random walk is $\alpha$, which models the rate of activity preservation. More specifically, while large values of $\alpha$ leads to a slow activity decrease, low values of $\alpha$ yields a more pronounced activity decrease. The size of the set of active nodes was computed varying $\alpha$ within the range $ 0.99990 \leq \alpha \leq 0.99995$. In all networks, an increasing trend with $\alpha$ could be observed. Specially, a strong correlation occurred for both scale-free and geographical networks. Therefore, the fraction of active nodes can be straightforwardly adjusted by setting the value of the parameter $\alpha$.}
    \end{figure}

    \begin{figure}
        \begin{center}
            \includegraphics[width=1\textwidth]{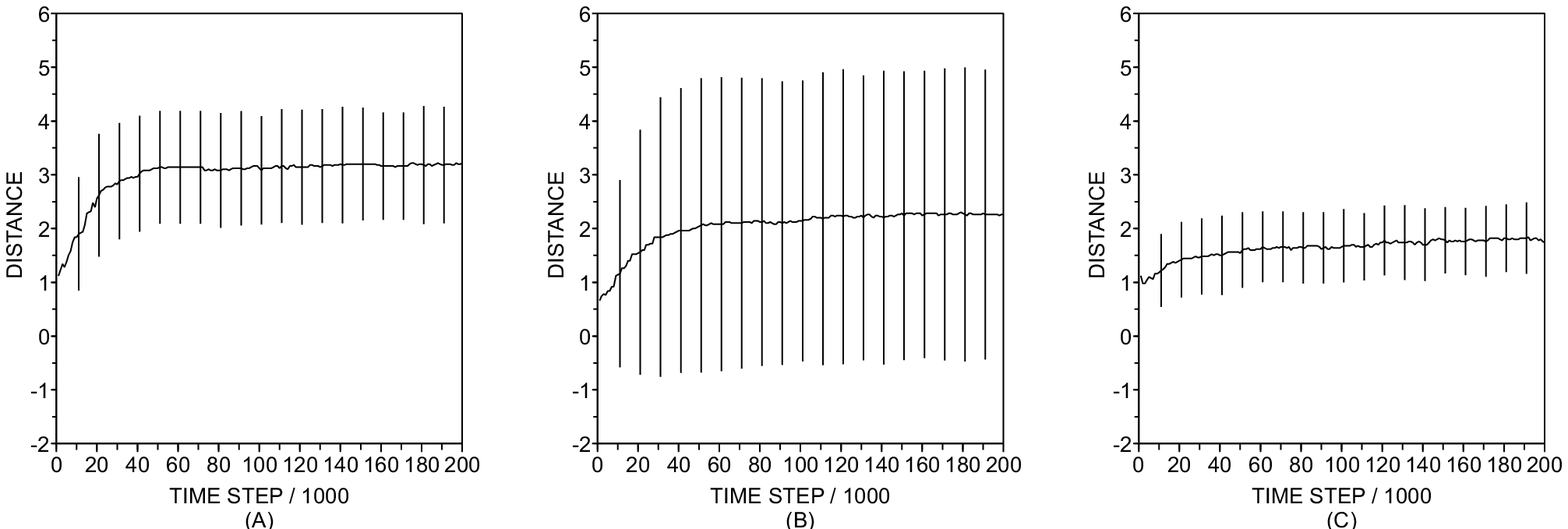}
        \end{center}
        \caption{\label{fig2.alf} Distance between the node starting the preferential random walk and the set of active active nodes at $t=t_f$, for (a) the random network; (b) the geographical network; and (c) the scale-free network. In all cases, the activation spreads along the nearest neighbors.}
    \end{figure}

    \begin{figure}
        \begin{center}
            \includegraphics[width=1\textwidth]{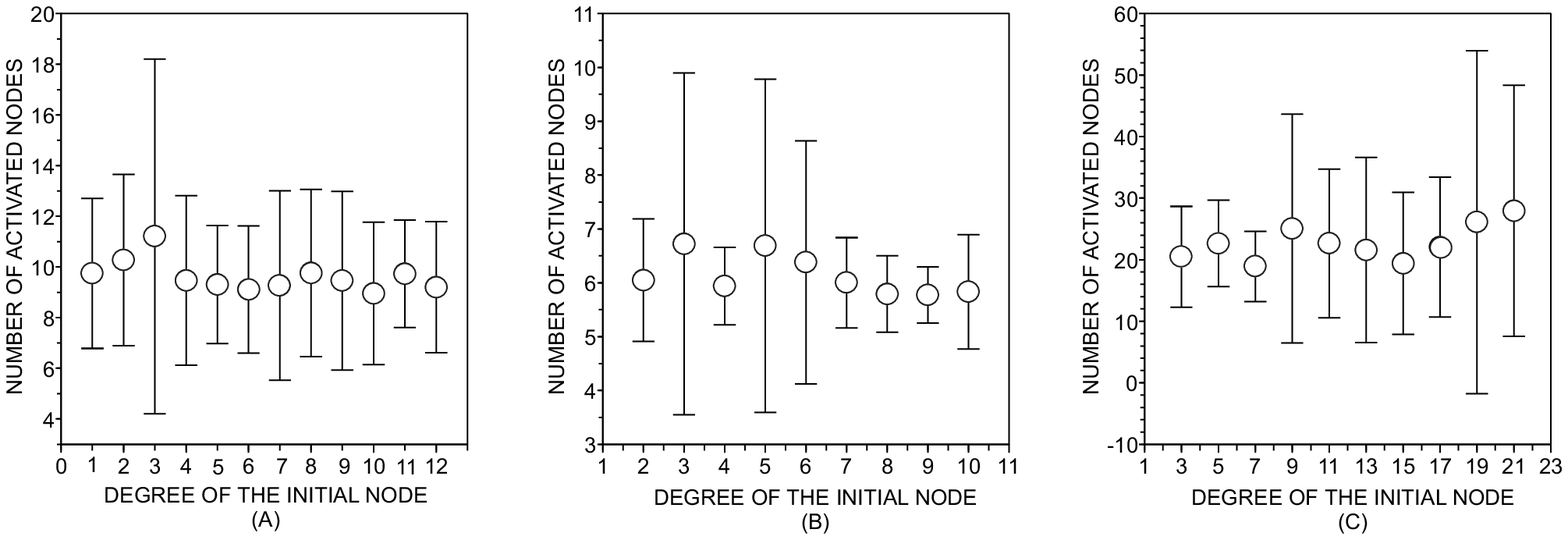}
        \end{center}
        \caption{\label{fig5} Size of the set of activated nodes and its relationship with the degree of nodes starting the preferential random walk  for (a) the random network; (b) the geographical networks; and for (c) scale-free network. In all cases, localization had weak impact on spreading. The influence of initial position of the particle was verified by starting the random walk in nodes with distinct degree. Then, the number of activated nodes at $t > t_f$ was computed for each class of starting nodes. Given the degree distributions in Figure \ref{fig4}, the set of degrees selected to start the walk was: $D_{r} = \{1,2,3,4,5,6,7,8,9,10,11,12 \}$, $D_{g} = \{2,3,4,5,6,7,8,9,10 \}$ and $D_{s} = \{3,5,7,9,11,13,15,17,19,21\}$ respectively for the random, geographical and scale-free networks. For the three network models, weak impact on spreading in the was observed. Interestingly, this result differs from the results reported in Ref.~\cite{kaiser}, which showed that a similar dynamics of activated nodes depends on where the initial activation was localized, especially for hierarchical modular networks.}
    \end{figure}

    \begin{figure}
        \begin{center}
            \includegraphics[width=1\textwidth]{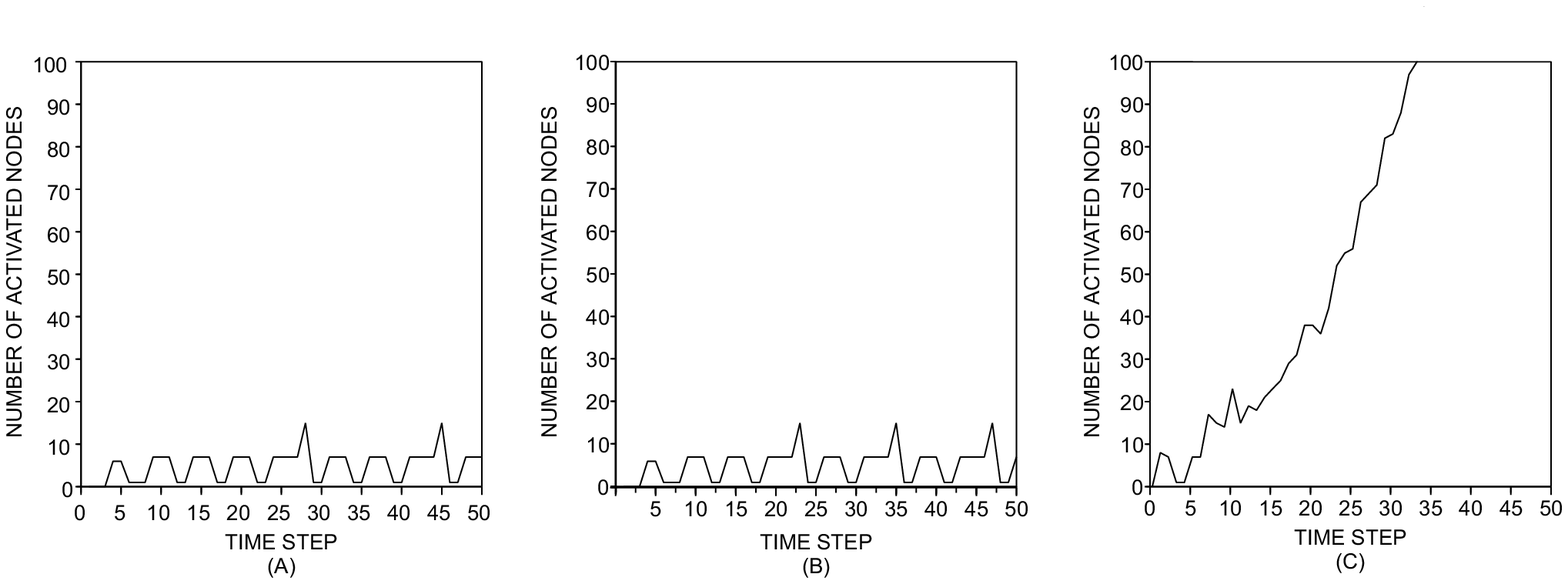}
        \end{center}
        \caption{\label{} Dynamics of the set of activated nodes  (i.e., nodes with $T_i > \tau_c$) submitted to the Integrate-and-Fire dynamics for the geographical network when (a) $\alpha = 0.8050$, (b) $\alpha = 0.8100$ and $\alpha = 0.81125$. While in (a) in (b) the activity is restrained, in (c) it spreads all over the network. }
    \end{figure}

    \begin{figure}
        \begin{center}
            \includegraphics[width=1\textwidth]{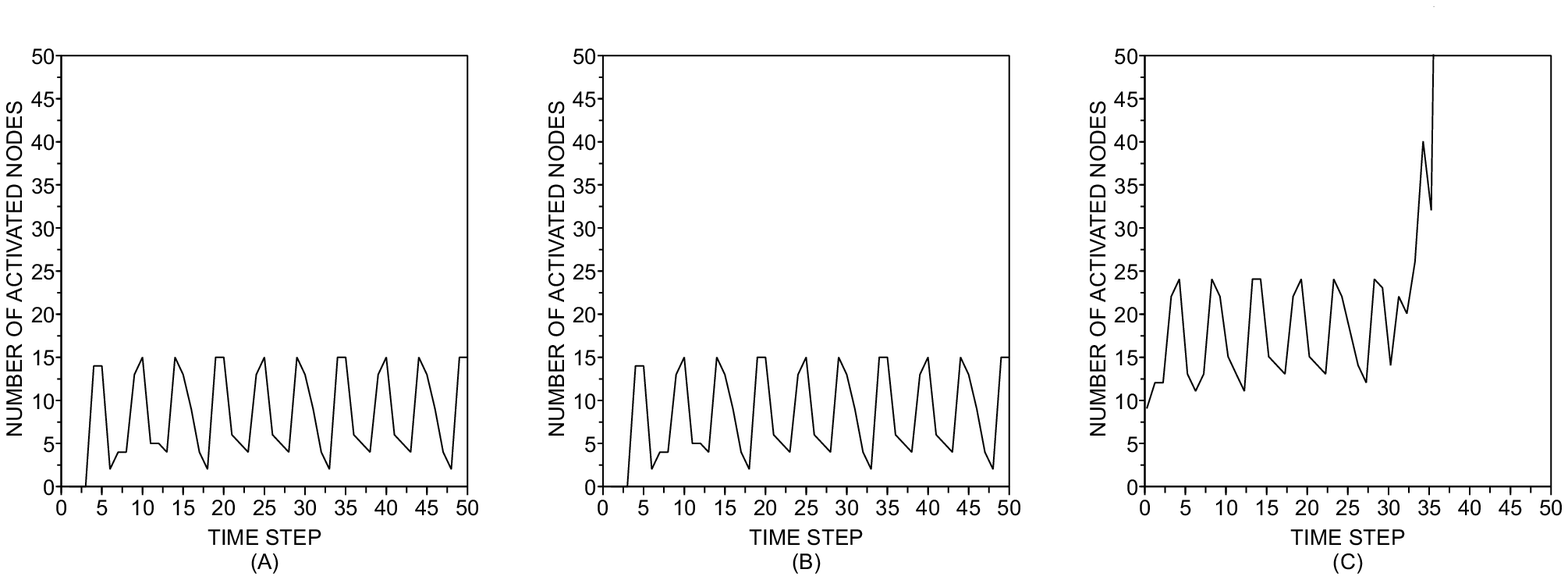}
        \end{center}
        \caption{\label{} Dynamics of the set of activated nodes (i.e., nodes with $T_i > \tau_c$) submitted to the Integrate-and-Fire dynamics for the BA network when (a) $\alpha = 0.7440$, (b) $\alpha = 0.7480$ and $\alpha = 0.7452$. While in (a) in (b) the activity is restrained, in (c) it spreads all over the network.}
    \end{figure}

\end{document}